\documentclass{aa}
\usepackage{graphicx}
\usepackage{txfonts}
\begin{document}
   \title{A Field Programmable Gate Array Spectrometer for Radio Astronomy}

   \subtitle{First Light at the Effelsberg 100-m Telescope}

   \author{S. Stanko
          \inst{1}
          \and
          B. Klein\inst{2}
          \and
          J. Kerp\inst{1}
          }

   \offprints{S. Stanko}

   \institute{Radioastronomisches Institut der Universit\"at Bonn,
              Auf dem H\"ugel 71, D-53121 Bonn\\
              \email{stanko@astro.uni-bonn.de}
              \email{jkerp@astro.uni-bonn.de}
         \and
              Max-Planck-Institut f\"ur Radioastronomie,
              Auf dem H\"ugel 69, D-53121 Bonn\\
              \email{bklein@mpifr-bonn.mpg.de}
             }

   \date{Received 22/10/2004; accepted 22/02/2005}

   \abstract{ We describe the technological concept and the
     first--light results of a 1024-channel spectrometer based on 
     field programmable gate array (FPGA) hardware.  This spectrometer
     is the prototype for the seven beam L-band receiver to be
     installed at the Effelsberg 100-m telescope in autumn 2005.
     Using ``of-the-shelf'' hardware and software products, we
     designed and constructed an extremely flexible
     Fast-Fourier-Transform (FFT) spectrometer with unprecedented
     sensitivity and dynamic range, which can be considered
     prototypical for spectrometer development in future radio
     astronomy.

   }

   \maketitle
%
%________________________________________________________________

\section{Introduction}
%-------------------------------------------------------------------------------

A wealth of information on the physical conditions of objects in the
universe can be gathered using spectrometers.  In the radio regime we
can roughly identify three basic types of spectrometers:
autocorrelators, acusto-optical spectrometers and filter banks.  Today
these spectrometers offer a useable bandwidth from a few kHz up to
2\,GHz with a few thousand spectral channels, which are capable of
resolving narrow spectral lines of masers and the thermal line
emission of gaseous clouds.

Over the past decade radio astronomy has been in a transitional phase
from single to multifeed arrays in continuum (bolometers) as well as
in heterodyne receivers (spectroscopy). Multifeed receivers enable very
sufficient surveys to large sky areas and significantly improve the
signal-to-noise ratio using the on-the-fly observing mode.  The huge
impact of a multifeed receiver on \ion{H}{i} and pulsar astronomy
(Manchester et al. \cite{manchester}) can be well illustrated by the
Parkes 21-cm multifeed system (Staveley-Smith \cite{staveley}).
Unfortunately, the backends do not develop at the same pace as the
receiver technology. Neither the number of spectral channels nor the
number of spectrometers nor the bandwidths have increased
significantly.

FPGA-chips might offer a solution for this  ``backend bottle neck''.
Using megagate chips, FFT- or polyphase filter bank algorithms can be
implemented today with spectral channel numbers ranging between 16,000 to
32,000.  The huge advantages of FPGAs are low price, wide
availablity (both based on the large commercial interest) and the
``plug-and-play'' installation as commercial personal computer (PC)
- PCIbus cards, which offer a high reliability.
Moreover, the sub-mm observatories at high altitudes have severe
difficulties to provide the receivers and supplementary power aggregates,
especially for digital autocorrelators that need several kW.
FPGA-spectrometers consume only a few W, so heat transfer via
coolers is unnecessary.
The low power consumption and compact size of FPGA-spectrometers
are attributes that
recommends this type of backend for use on spacecraft and satellites.

In this paper we present our design approach toward developing an
FPGA-spectrometer for radio astronomy. The technical concept is
presented in Sect.\,2 of this {\em research note\/}, followed by the
basic implementation.  {\em First
  light\/} observations of the \ion{H}{i} and OH maser line emission
at the Effelsberg 100-m Telescope are presented in Sect.\,3.

%__________________________________________________________________

\section{Techniques}

%                                     Two column figure (place early!)
%______________________________________________ Gamma_1 (lg rho, lg e)
   \begin{figure*}
   \centering
   \includegraphics[width=17cm]{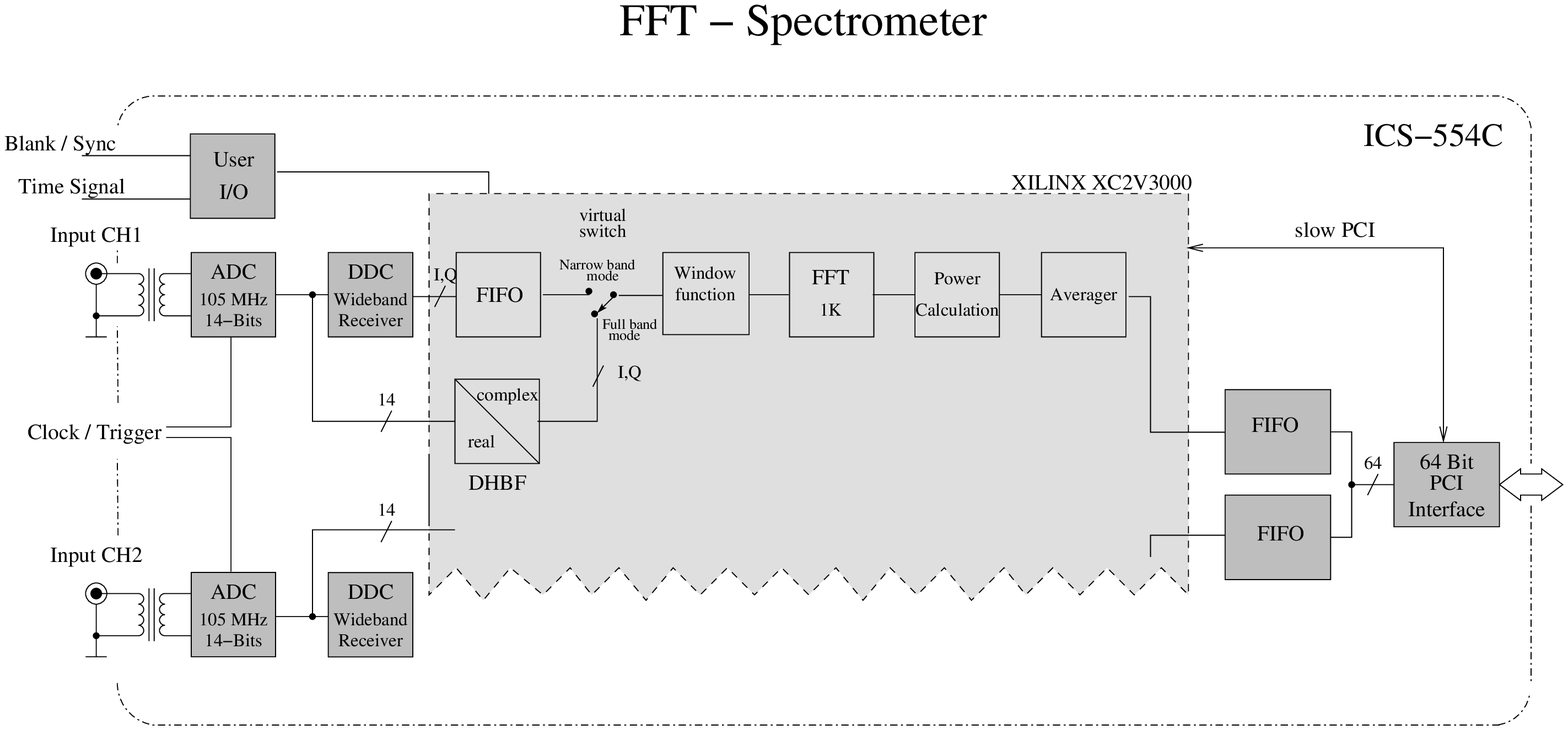}
   \caption{Sketch of the ICS\,554C device and the arrangement of
     Intellectual Property (IP)-cores to constitute an
     FFT-spectrometer on the field programmable gate array
       (FPGA).  The two signals from the 21-cm receiver enter the
     device on the left-hand side via the analog to digital
       converters (ADCs).  The optional  digital down converters
       (DDCs) reduce the bandwidth to values from 10\,MHz to a few
     kHz.  Within the FPGA, we implement a distributed halfband
       filter (DHBF), a window-function (look-up table) followed by
     the FFT.  Both the DDC and the DHBF produce complex data values
     (I: in-phase, Q: quadrature). The pipeline FFT output is
     transferred into the power-spectrum calculator which sums the
     spectra in an averager.  The spectra are stored on the PC via the
     first-in first-out pipelines (FIFOs). }
              \label{blockdiagram}%
    \end{figure*}
%
%______________________________________________________________
   \begin{figure}
   \centering
   \includegraphics[width=8cm]{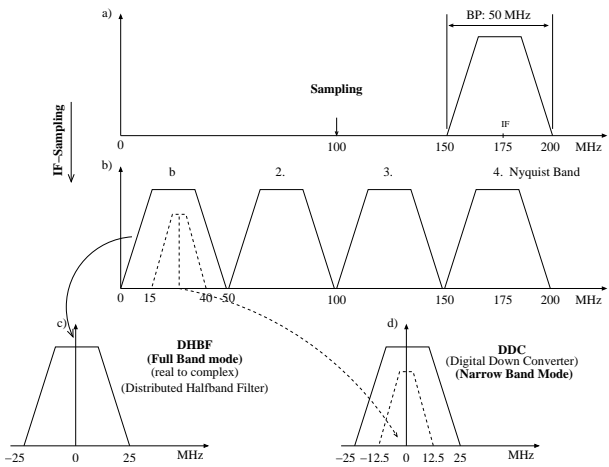}
      \caption{Principle of downconverting in Full Band Mode and
        downconverting with additional filtering, mixing and
        decimation in Narrow Band Mode.  The output frequency of
          the receiver IF band is 150 to 200\,MHz (a). Because of the
          sampling frequency of 100\,MHz the 4th Nyquist Band is
          sampled (b). In DHBF mode the whole band
          with center frequency of 25\,MHz ist converted to 0\,MHz
          (c). In Narrow Band mode (using the DDCs) a part of the band
          (dashed in the example) with a user selectable center
          frequency is converted to 0\,MHz.  }
         \label{downconverter}
\end{figure}

\subsection{Concept}
%-------------------------------------------------------------------------------

One of the most important fundamentals for correctly sampling analog
signals is the concept of aliasing.
The famous Nyquist criterion states that data recorded at a rate of
$F_s$ samples per second can effectively represent a signal of
bandwidth up to $F_s/2$ Hz. Sampling signals with greater bandwidth
produces aliasing: signal components at frequencies greater than $F_s/2$
are folded, or aliased, back into the $F_s/2$ band.

The frequency translation aspect of aliasing  can be exploited
by using a technique called \emph{undersampling},
\emph{band-pass sampling}, \emph{harmonic sampling} or
\emph{Super-Nyquist sampling}.  To understand this technique, one must carefully 
consider the definition of the Nyquist constraint. Sampling
a signal of bandwidth $F_s/2$ requires a minimum sample rate
$F_s$, but this $F_s/2$ bandwidth can theoretically be located
anywhere in the frequency spectrum (e.g. $n F_s$ to $(n + 1/2)\,F_s$,
$n=0,1,2,...$), not simply from DC to $F_s/2$ like baseband signals.
The aliasing action, like a digital mixer, can be used to translate an
intermediate frequency (IF) down to the baseband (see
Fig.\,\ref{downconverter}).  Signals $s$ in the bands $n F_s < s < (n +
1/2)\,F_s$ will be translated down intact; the translation of signals in the bands $(n
- 1/2)\,F_s < s < n F_s$ will be ``mirrored'' in frequency,
an effect corrected in the PC after the measurement.

Using undersampling techniques (Groshong \& Ruscak \cite{groshong})
enables direkt IF-band sampling of the receiver outputs without any
additional analog baseband mixing.  Only one (antialiasing) bandpass
filter in front of the analog-digital converter (ADC) is necessary to
select the Nyquist band and avoids aliasing from other frequency
bands.  This concept saves costs and, because analog mixers are highly
sensitive to input levels, increases the signal quality. If not
carefully adjusted, the input signal can cause considerable harmonic
distortion, which, in addition varies as a function of temperature.

The system is configured to process two channels in
parallel, each channel  with a 50\,MHz bandwidth.
It is based on a data acquisition card ICS554C (ICS 
\cite{ics}), consisting of two 14-bit ADCs running at 100\,MHz
sample rate, a Xilinx FPGA, two FIFO-buffers and a PCIbus interface
(Fig.\,\ref{blockdiagram}).

The on-board Xilinx XC2V3000 FGPA is capable of processing the signal
from both ADCs in parallel to produce 2\,$\times$\,1024 channel
spectra.  The digital data streams produced by both ADCs are connected
to the FPGA and piped in through separated Digital Down Converters
(DDC) chips which are installed to determine the bandwidth (increase
the spectral resolution) of the signal and to select sub-bands.  Any
necessary mixing and filtering is done digitally in these DDCs. No
analog elements are used and reconfigurating the sub-band and
bandwidth is easily done via the PC.  The DDCs are GC4016 quad
receivers (Graychip \cite{gray}), each one containing four identical
down conversion circuits. The input data stream can be received in
four separate sub-bands, each with up to 2.5\,MHz bandwidth. By
combining all of these, a maximum bandwidth of 10\,MHz can be
processed, corresponding to a spectral resolution of about 10\,kHz per
spectral channel. The best resolution of $\sim\!20$\,Hz per channel can be
achieved by reducing the minimum bandwidth to 20\,kHz.

\subsection{Implementation}
%-------------------------------------------------------------------------------

Individual cores were purchased from RF-Engines to implement the processing pipeline shown in Fig.\,\ref{blockdiagram}. This straightforward
implementation and concept makes it
possible to obtain debugging information at any stage of the
pipeline.

The ``raw'' data from the ADCs, as well as the data processed by the
DDCs, are routed into the FPGA in parallel. A virtual switch has been
implemented in order to select data either as {\em Full Band} (50\,MHz
bandwidth) from ADC (also referred to as the Distributed Halfband
Filter, DHBF) or {\em Narrow Band} (10\,MHz -- 20\,kHz bandwidth) from
the DDCs.  When using the DHBF, the bit stream is mixed with sine and
cosine waves to shift the signal, centered symmetrically about the IF
of half the bandwidth, to a signal centered at 0\,Hz (see
Fig.\,\ref{downconverter} (b--c)).   In contrast to  real-only digitization this technique yields a full
Nyquist sampling, and the
following pipeline FFT works with the highest efficiency on the
complex data.

An additional
FIFO had to be added to deal with data by the DDCs. This is because the pipeline always transfers
data of exactly 2048 samples, but the data produced by the DDCs are not
delivered continuously.

A window function is applied to the data first (Kaiser
\cite{kaiser}).  This function is freely programmable via the PC without re-loading the FPGA design by means of
``slow'' PCIbus connection.

The central element of the whole processing pipeline is the Vectis
HiSpeed 1024-point radix--2 FFT algorithm developed by RF-Engines.
This pipelined FFT-core works up to 100\,MHz, enabling a continuous
Fourier Transformation of the input signals.  The output consists of
27-bit interleaved real and imaginary signed data, which are fed into
the power spectrum calculator and the following averager.

There are two possible  communication channels between
the FPGA and the PC: on the one hand  a ``slow'' 8 bit PCI control
bus, on the other hand the ICS554C card with two outgoing hardware
FIFOs, which can transfer data via a 64--bit, 64\,MHz PCI bus. This
gives a maximum PCI burst data transfer speed of up to
528\,MB\,${\rm s}^{-1}$.

The control bus is used to set window, integration time
(number of samples) and observing mode, as well as a readout of data
ready and status information. The FIFOs are used to transfer
the Fourier transformed and pre-integrated spectra. The high speed
connection is essential to keep the sub-samples integration time low,
e.g.
for radio frequency interference  (RFI) investigations or pulsar
measurements.

PC software was developed that is capable of setting control values
for the window, integration time and the configuration of the DDCs.
The PC is responsible for triggering the first sample and then wait
for a data ready signal. The PC software then reads back the
integrated spectra from 2 channels using the hardware FIFOs in FPGA
and DMA mode. The data are stored as ASCII-files with
an optional header.

\subsection{Stability Tests}
%______________________________________________________________
   \begin{figure}
   \centering
   \includegraphics[width=8cm]{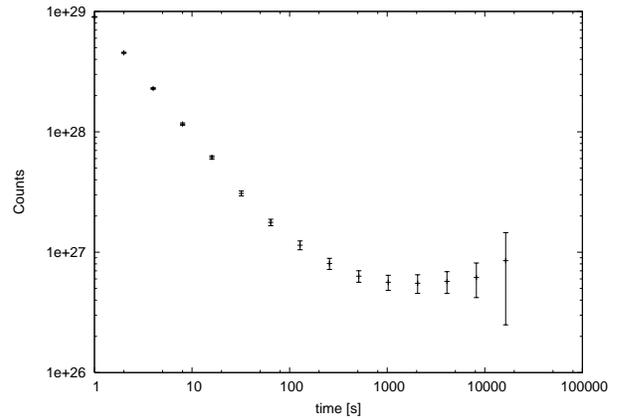}
      \caption{Allan-plot of the stability of the FPGA-spectrometer.
              }
         \label{allanplot}
\end{figure}
%
%______________________________________________________________
   \begin{figure}
   \centering
   \includegraphics[width=8cm]{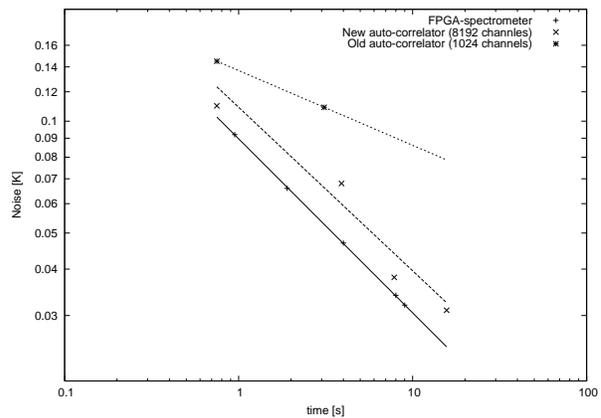}
      \caption{Effelsberg 100-m RMS-limit temperature as a function of
        integration time of the 1024 and 8192
        spectral channel autocorrelators
with the FPGA--spectrometer. The autocorrelators are the standard
  spectrometers at the Effelsberg 100-m telescope. The 1024 channel
  autocorrelator samples 1-bit, the 8192 channel autocorrelator
  2-bit/3-level. 
              }
         \label{noiseplot}
\end{figure}
In order to quantify the stability of the spectrometer, we examined
its performance when connected to a wideband noise generator. The
Allan plot (Schieder \cite{schieder}) generated by this exercise
implies us stability times $\approx 1000$ seconds (see
Fig.\,\ref{allanplot}). Compared to other spectrometers this is
excellent and highly sufficient for astronomical applications.

The first quantitative evaluation of the IAU standard position S7  \ion{H}{i}-line spectra (Kalberla et al. \cite{kalb82})
suggested that the new FPGA-spectrometer using samplers with
  14-bit dynamics is a factor of about 
  $\pi/2$  more sensitive than
the 1024-channel autocorrelator,  which uses 1-bit sampling.  A few days
later in a second observing run,
we were able to observe S7 with a cold 21-cm receiver. We compared the
FPGA-spectrometer data with the 1024- and 8192-autocorrelator 
(which uses 2-bit/3-level sampling) spectra observed in parallel.  Our
results are plotted in Fig.\,\ref{noiseplot}.

The sensitivity of the system, given by the standard radiometer formula 
\begin{equation}
\hfill\frac{\Delta T}{T_{\hbox{sys}}} = \frac{C}{\sqrt{B\,t_{{\rm int}}}},\hfill
\end{equation}
with $C=\frac{\pi}{2}$ for 1-bit autocorrelators,  $C=1.23$ for the
  2-bit/3-level autocorrelator, and $C \simeq 1$ for the FPGA-spectrometer
(Hagen \& Farley \cite{hagen}), clearly indicates superior sensitivity
of the FPGA-spectrometer. The fit gives us a ratio of 1.53 between
  the FPGA-spectrometer and the 1024 channel / 1-bit autocorrelator C
  values and a ratio of 1.21 between the FPGA-spectrometer and the
  8192 channel autocorrelator / 2-bit, 3-level.

\section{Observation}
%-------------------------------------------------------------------------------
%
%                                                One column figure
%----------------------------------------------------------- S_vib
   \begin{figure}
   \centering
   \includegraphics[width=8cm,totalheight=4cm]{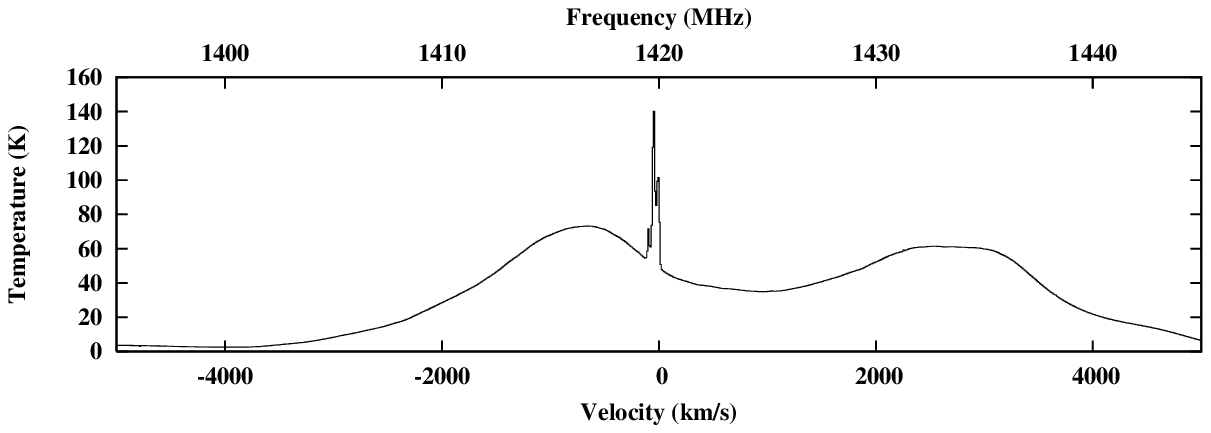}
   \includegraphics[width=8cm,totalheight=4cm]{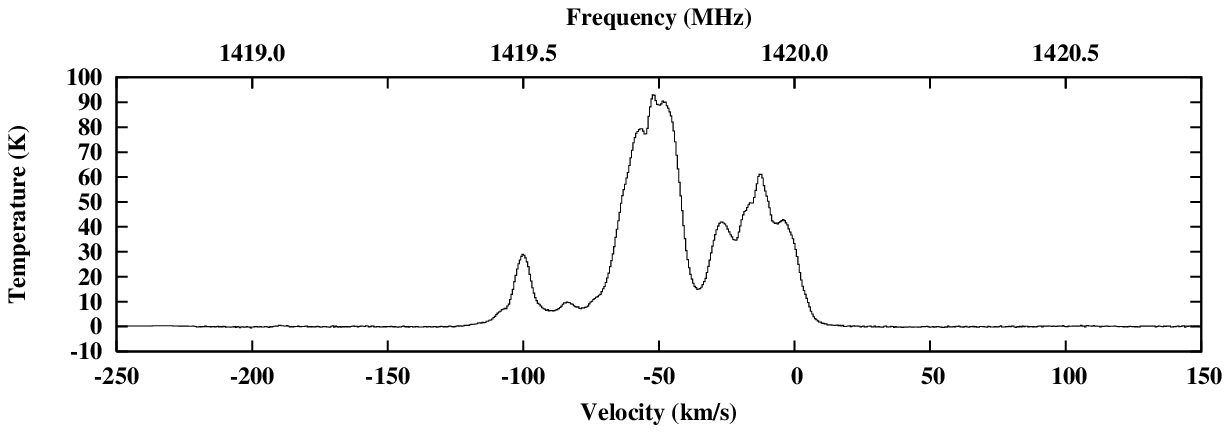}
      \caption{{\bf Top:} \ion{H}{i} 21-cm line spectrum of the
        standard calibration source S7 observed using the total
        bandwidth of 50\,MHz of the FPGA-spectrometer. The
        hardware bandpass was
not optimized to be flat across the whole frequency range.
{\bf Bottom:} {\em ON -- OFF\/} subtracted DDC \ion{H}{i} spectrum of the S7 source with 1 minute
of integration time. The  bandwidth is about 2\,MHz. 
              }
         \label{spectra}
   \end{figure}
%
%______________________________________________________________
   \begin{figure}
   \centering
   \includegraphics[width=8cm,totalheight=4cm]{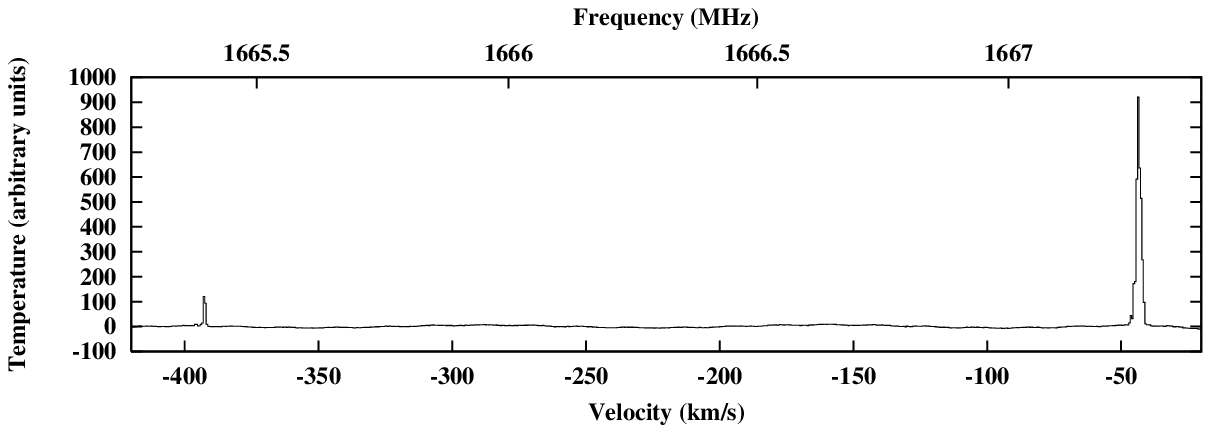}
   \includegraphics[width=8cm,totalheight=4cm]{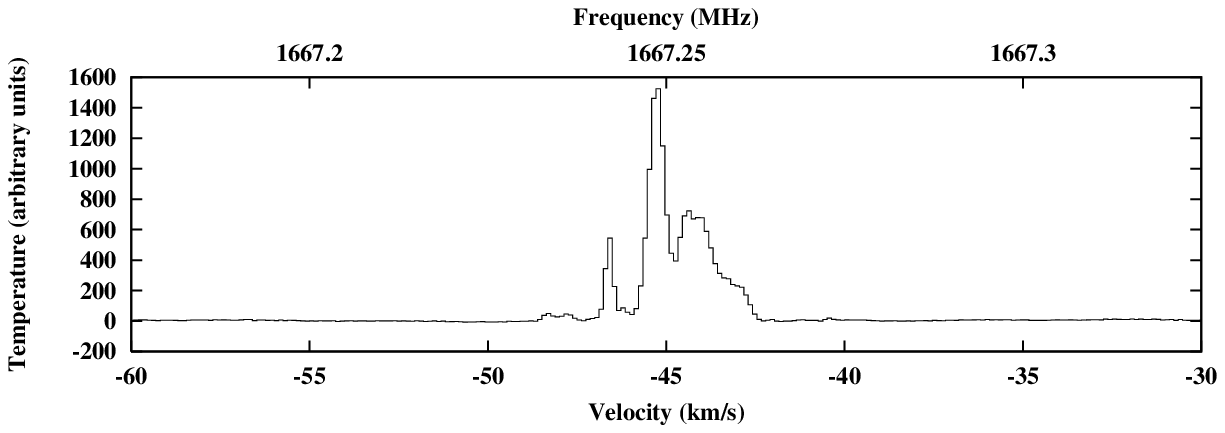}
      \caption{{\bf Top:} OH-maser emission of the W3OH star formation
        complex observed with a bandwidth of about 2\,MHz. The
        intensity contrast between the main emission line at 1667\,MHz
        (right) and the satellite line at 1665\,MHz (left) is an
        order of magnitude. {\bf Bottom:} Resolved line emission of
        the OH maser emission at 1667\,MHz (one circular polarisation
        channel) with a bandwidth of 0.2\,MHz using the DDCs. Please
        note that the temperature scale is not calibrated in contrast to
        the \ion{H}{i} spectra! 
          }
         \label{w3oh}
   \end{figure}

   The {\em first light} observation using our FPGA-spectrometer
   occured on 24 August 2004. To study the quality of the \ion{H}{i}
   21-cm line spectra we chose the northern sky standard calibration
   source S7.  For a first look, we used the whole band (50\,MHz) of
   the FPGA-spectrometer.  Fig.\,\ref{spectra} (top) shows the
   recorded spectrum of the standard calibration source.  We intended
   to use a frequency switching mode to determine the baseline of the
   spectrometer as accurately as possible.  This was done by
   integrating for 30\,seconds at the center-frequency (ON spectrum),
   shifting the frequency by tuning the receiver by 6\,MHz, and
   integrating for an additional 30\,seconds (OFF spectrum).  To
   normalize the line emission detected by the FPGA-spectrometer, we
   used the S7-spectra observed in parallel with the autocorrelator
   spectrometer operated and, implementing standard \ion{H}{i} data
   reduction procedures (Peter Kalberla, priv. comm.)  determined the
   system temperature and evaluated the quality of the S7-calibration
   observation.  The velocity resolution of the 50\,MHz FPGA data
   corresponds to 10.32\,$\mathrm{km\,s^{-1}}$.  We computed the
   difference of the ON and OFF spectra and performed a relative
   calibration using a part of the baseline which is free of
   astronomical line emission.
   
   Since the velocity channel width of 10.32\,$\mathrm{km\,s^{-1}}$ is
   much too coarse for galactic \ion{H}{i} studies, we switched to the
   DDC mode, setting the initial bandwidth of 2\,MHz.  88\% of this
   bandwidth is scientifically usable because of the digital bandpass
   of the DDCs, which suppresses the band edges.  Following the same
   frequency switching procedure as described above, we observed the
   S7-line profile with a 0.5\,$\mathrm{km\,s^{-1}}$ velocity
   resolution (Fig.\,\ref{spectra} bottom).  The baseline subtraction
   was performed as described above.

To evaluate the dynamic-range of the FPGA-spec\-tro\-meter, we
observed the OH maser lines of the W3OH-star forming region (Norris \&
Booth \cite{norris}).  Using a 2\,MHz bandwidth, we simultaneously
observed 120\,K line emission at 1667\,MHz  and about an order
of magnitude stronger line emission of the maser at 1665\,MHz.
Fig.\,\ref{w3oh} (top) shows these data without any baseline subtraction.
All DDC spectra reveal a sinusodial baseline modulation, which is
  very stable in time.
Accordingly we always obtain an accurate
flat baseline after a simple ON-OFF subtraction as described above.
Using the DDCs to increase the velocity resolution to
0.03\,$\mathrm{km\,s^{-1}}$, we can resolve an individual line complex
(Fig.\ref{w3oh} bottom).
Please note that the OH-spectra, in
contrast to the S7 \ion{H}{i}-spectra, are not normalized.  It is worth noting that we do
not need to change any attenuator in our setup when switching from the weak
       continuum S7 21-cm line observations to the OH-maser region.
This is necessary when using the autocorrelator spectrometer
with its limited dynamic range.

\section{Conclusion}
%-------------------------------------------------------------------------------
{\em First light\/} observations using an
FPGA-spectrometer have been performed at the Effelsberg 100-m Telescope. These observations
demonstrate that FPGA-spectrometers are capable of superseding current
standard spectrometers with their limitations in bandwidth, number
of spectral channels and dynamic range.
Moreover, the sensitivity is superior to autocorrelators
due to the larger number of ADC bits.
Considering the high redundancy and low costs offended by commercial products, FPGA-spectrometers might well be the future in radio astronomy spectroscopy.

\begin{acknowledgements}
  We would like to acknowledge the major contribution of Ingo Kr\"amer
  to overcome minor and major difficulties during the FPGA
  programming.  The excellent cores of RF-Engines ({\tt http://www.rfel.com}) allowed us to
  develop the FPGA-spectrometer on a time-schedule of only a few
  months.  We would also like to thank Alexander Kraus for the rapid
  scheduling of the test observing time at the 100-m Telescope and
  Mike Bird for carefully reading the manuscript.
\end{acknowledgements}

\end{document}